# Phase diagram of Rydberg-dressed atoms on two-leg triangular ladders

Pierre Fromholz,[1,*] Mikheil Tsitsishvili,[2,3,†] Matteo Votto,[2] Marcello Dalmonte,[2,3] Alexander Nersesyan,[2,4,5] and Titas Chanda[2,‡]

[1] *Department of Physics, University of Basel, Klingelbergstrasse 82, CH-4056 Basel, Switzerland*
[2] *The Abdus Salam International Centre for Theoretical Physics (ICTP), Strada Costiera 11, 34151 Trieste, Italy*
[3] *International School for Advanced Studies (SISSA), via Bonomea 265, 34136 Trieste, Italy*
[4] *The Andronikashvili Institute of Physics, 0177 Tbilisi, Georgia*
[5] *Ilia State University, 0162 Tbilisi, Georgia*

Dressed Rydberg atoms in optical lattices are a promising platform for the quantum simulation of intriguing phenomena emerging in strongly interacting systems. Relevant to such a setup, we investigate the phase diagram of hard-core bosons in a triangular ladder with next-to nearest-neighbor interaction along each leg and nearest-neighbors interactions without hopping between the legs. For weak interactions, Abelian bosonization predicts a spin density wave and a fully gapless Luttinger liquid phase. Such liquids transition to a 'spin-locked' cluster Luttinger liquid at strong interactions along each leg, as predicted by cluster bosonization. Interestingly, the competition with the zigzag interaction generates a charge density wave, a 'polarized holonic' phase, and a crystalline phase at the filling 2/5, that we address via semi-classical perturbative approach. Exact diagonalization and density matrix renormalization group simulations confirm the predictions and further characterize the phases and their transitions.

*Introduction.* Over the last ten years, ensembles of ground-state atoms laser-coupled to Rydberg states in optical lattices have shown outstanding aptitudes for realizing strongly interacting quantum dynamics under controlled and tunable experimental conditions [1–7]. For example, a weak coupling between the ground state and the Rydberg states [8–11] (the Rydberg-dressing) generates an effective long-range two-body interaction with a short-range plateau (the soft-shoulder potential [12]) and a power-law tail (van der Waals). The range of the plateau may reach a few microns such that probing the single-site level is easier than conventional cold atoms systems operating in Hubbard-like regimes. For a single scalar bosonic field in 1D, such an interaction induces exotic critical behaviour such as cluster Luttinger liquids (CLL) where the Luttinger theorem is inapplicable [13] and supersymmetric (SUSY) quantum critical points [14]. In 2D systems, it is associated with anomalous dynamics and glassy behavior [15, 16].

Constrained dynamics is instead relatively unexplored in mixed dimensional settings, in particular ladder systems. Such geometries are particularly interesting, as they have been recently realized in experiments [17]. In a first work along this direction, we considered square ladder lattices for dressed Rydberg atoms at filling $\nu = 2/5$, with the two legs interacting, and without inter-leg hopping [18]. Such geometry effectively couples two SUSY theories at a fine-tuned point separating a Tomonaga Luttinger liquid (TLL) and a CLL. All phases and transitions involved in the phase diagram surprisingly appear as soon as the two chains interact. Both the TLL and CLL are 'spin-locked' and the critical point extends into a Gaussian phase transition line, a possibly SUSY conformal transition, and a partially gapped phase in between.

In this work, we extend this exploration to triangular lattice geometries. Such a setting is the minimal geometry where geometric frustration, interaction clustering, and kinetic dynamics can compete. Our main result is the phase diagram of the system that is schematically depicted in Fig. 1: it is considerably richer than - and fundamentally different from - the square lattice case. The ability to manipulate the coupling of the two SUSY conformal field theories by *e.g.,* increasing the range of the inter-leg interaction would be useful to investigate SUSY phenomenology [19, 20]. By using a triangular geometry, we show that as soon as the range is extended, only a single, potentially SUSY, transition exists between the 'spin-locked' TLL and CLL. In contrast, we show the existence of three phases exclusive to strong coupling, establishing the setup as a promising platform to study commensurability in extended Hubbard models.

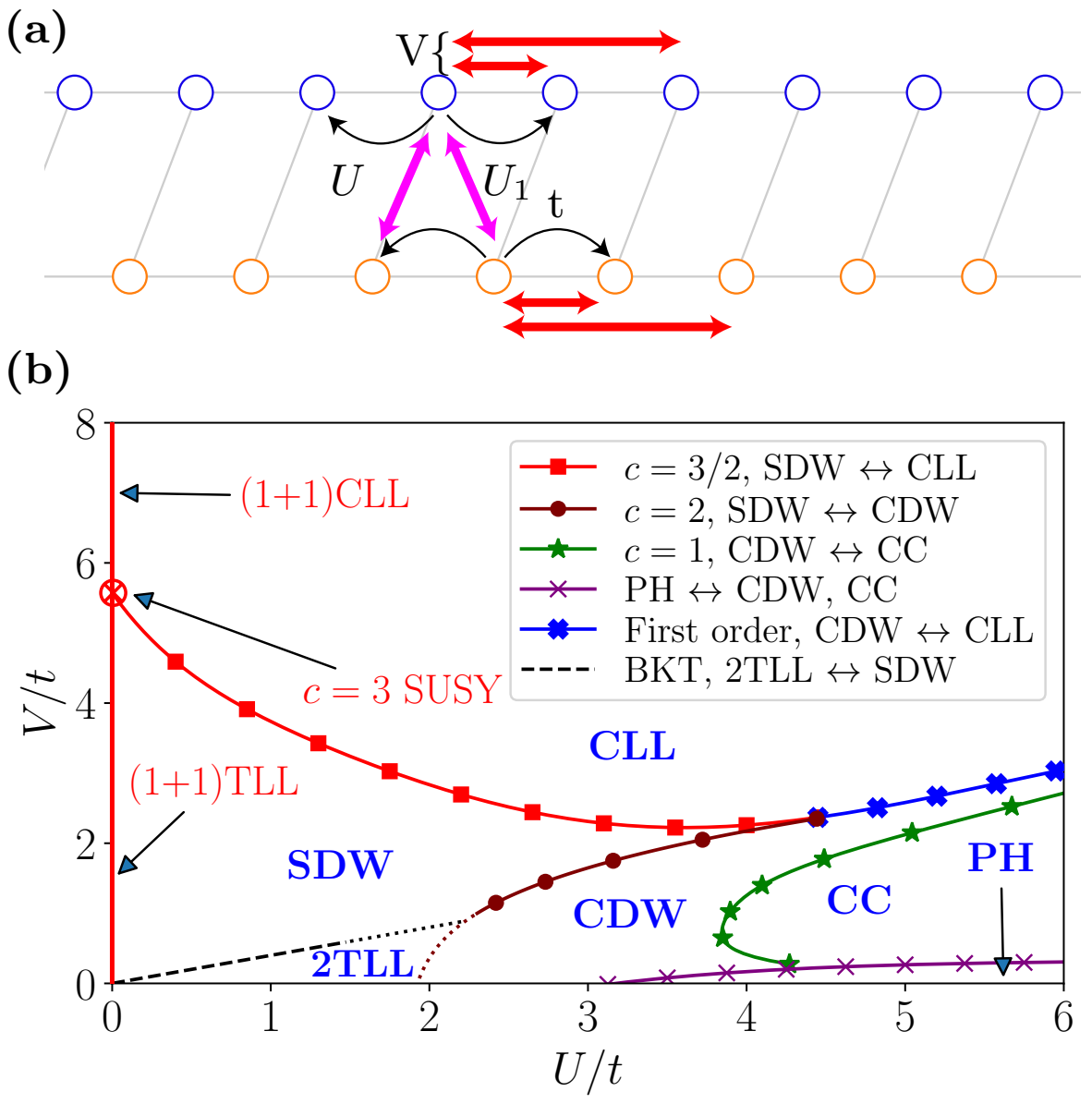

FIG. 1. (Color online.) **(a)** The system of Rydberg-dressed hardcore bosons on a two-leg triangular lattice with intra-leg soft-shoulder interaction ($V$) of range 2, anisotropic inter-leg interactions ($U, U_1$), and intra-chain nearest-neighbor hopping ($t$). **(b)** The reach landscape of the phases of the triangular ladder system for a filling of 2/5 with $U_1 = U$. For $U/t > 0$, the phase diagram consists of $c = 2$ Tomonaga-Luttinger liquid (2TLL), $c = 1$ spin density wave (SDW), $c = 1$ charge density wave (CDW), $c = 1$ cluster Luttinger liquid (CLL), polarized holonic (PH) phases, and fully-gapped cluster crystal (CC) (see text) as marked in the figure. Different lines correspond to different phase transitions as depicted in the legend. At $U = 0$ (red line) the two legs are independent and display $c = 1 + 1$ TLL and $c = 1 + 1$ CLL phases separated by $c = 3$ supersymmetric point (red crossed-dot).

*Triangular ladder.* We study a triangular ladder of spinless hard-core bosons at $\nu = 2/5$ filling (on average two filled sites for every five sites in each chain, see Fig. 1**(a)**). Such a filling supports clustering in the classical limit. The Hamiltonian of the system reads:

$$H = -t\sum_{i,\ell}\left(b^{\dagger}_{i,\ell}b_{i+1,\ell} + \text{H.c.}\right) + U\sum_{i} n_{i,+}n_{i,-} + U_1\sum_{i} n_{i,+}n_{i+1,-} + V\sum_{i,\ell}\sum_{j=1}^{r_C} n_{i,\ell}n_{i+j,\ell}. \quad (1)$$

where $b^{\dagger}_{i,\ell}$ is the creation operator for hard-core bosons on site $i$ of chain $\ell = \pm$ and $n_{i,\ell} = b^{\dagger}_{i,\ell}b_{i,\ell}$. $t$ is the intra-chain hopping, $U$ and $U_1$ are the anisotropic nearest-neighbor inter-chain interaction amplitude, $V$ is the intra-chain interaction amplitude of range $r_C \in \mathbb{N}$. As there is no inter-chain tunneling, the U(1) charge of each leg is conserved separately such that the Hamiltonian Eq. (1) is U(1)×U(1) symmetric. Densities are tuned by the loading scheme. For simplicity, we take $r_C = 2$ and $U_1 = U$ unless otherwise stated.

We now briefly summarize the phase diagram at weak-coupling, that is obtained using both weak-coupling bosonization, cluster bosonization. These approaches follows closely previous works [13, 14, 18]. We thus refer to to the supplementary material [21] for the analytical details of the approach, and only present its results below.

The red line ($U = 0$) in Fig. 1**(b)** corresponds to two decoupled chains [13, 14]. The phase diagram of a single chain displays a TLL for $0 < V/t \lesssim 5.7$, and a CLL for $V/t \gtrsim 5.7$. Both the TLL and the CLL have a central charge $c = 1$. They are separated by a SUSY conformal phase transition point of central charge $c = 3/2$ at $V/t \simeq 5.7$ identified numerically [14].

By adding the interaction ($U$) between two of these chains, we obtain the rest of the phase diagram as shown in Fig. 1**(b)**. We map the leg index to a SU(2) spin-1/2 degree of freedom to use the Hubbard model formalism and vocabulary [12, 22, 23]. When $U > 0$, the phase diagram displays a Spin Density Wave (SDW) TLL with $c = 1$ and a fully gapless TLL with $c = 2$ (2TLL) at weak coupling when $U/t \leq 5V/2t \ll 1$ and $5V/2t \leq U/t \ll 1$ respectively. These two are separated by a Berezinskii-Kosterlitz-Thouless (BKT) transition.

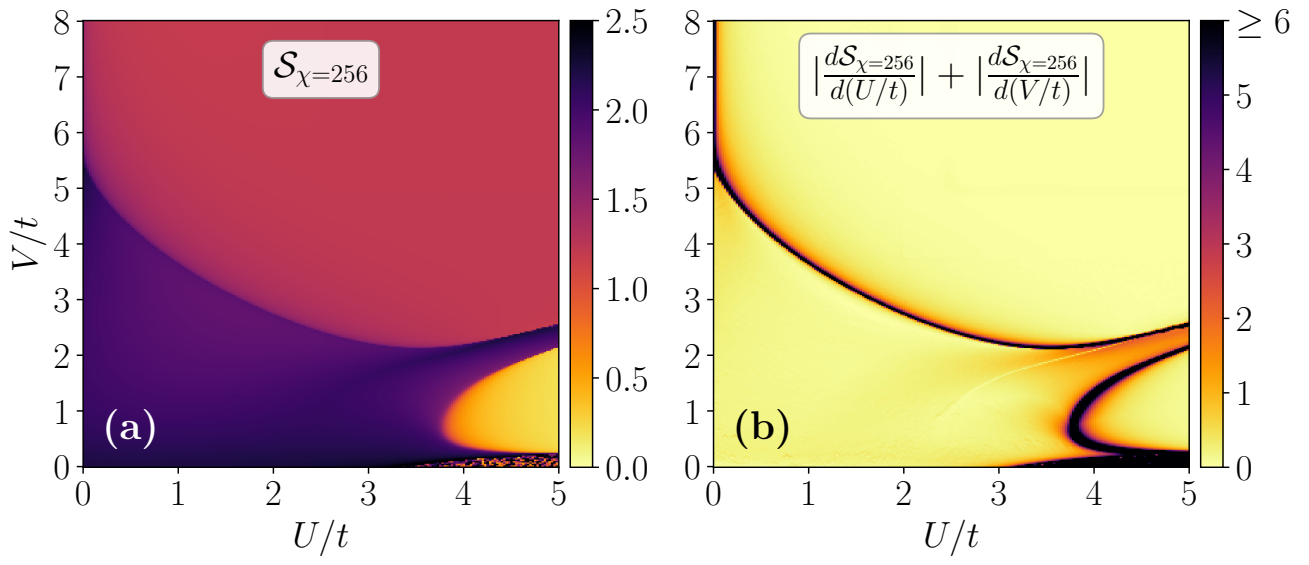

FIG. 2. (Color online.) **(a)** The entanglement entropy $\mathcal{S}$ and **(b)** the sum of the absolute values of derivatives with respect to the system parameters $U/t$ and $V/t$ in the $(U/t, V/t)$-plane. The quantities are computed by the iDMRG algorithm with bond dimension $\chi = 256$. The BKT and 2TLL ↔ CDW transitions are too narrow to stand out when $\chi = 256$.

For strong shoulder potential ($V/t \gtrsim 5.7$), we show that the phase is a spin-locked CLL with $c = 1$ for $U \neq 0$. We predict the separation of the SDW and the CLL by a single phase transition line of central charge $c = 3/2$ extending from the $c = 3$ SUSY critical point. The value of this central charge hints that this phase transition line might itself be supersymmetric. This behavior contrasts with the square ladder geometry [18] where the SDW and CLL are separated by two transitions and one Luttinger liquid phase instead. The separation by a single transition occurs as soon as $U, U_1 > 0$. Furthermore, the phase diagram Fig. 1**(b)** near the SUSY point is qualitatively unchanged for any $U_1/U > 0$ (c.f. [21]) justifying taking $U_1 = U$ for simplicity.

*Numerical simulations.* We confirm these predictions and explore the large $U/t$ regime using both exact diagonalization (ED) [24] for a system of 10 sites with periodic boundary condition and infinite density matrix renormalization group (iDMRG) techniques [25–32].

We first characterize the system using iDMRG to effectively access the system directly at the thermodynamic limit. The set $\chi$-th value of the bond dimension of the matrix-product state ansatz for the iDMRG simulation introduces a maximal length-scale (sometimes referred to as correlation length [33]) $\xi_\chi$ beyond which any correlation decays exponentially [34–36]. By computing the bipartite von Neumann entanglement entropy $S_\chi$, we obtain the phase diagram in Fig. 2.

Besides contours compatible with the weak $U/t$ predictions (when visible), the simulation predicts three additional phases at larger $U/t$. Anticipating their identification, we find a Charge Density Wave (CDW) TLL phase at intermediate $U/t$. When $U/t \gg 1$ and $V/t \gtrsim 0.34$, we find a fully gapped cluster crystal (CC). When $U/t \gtrsim 3.1$

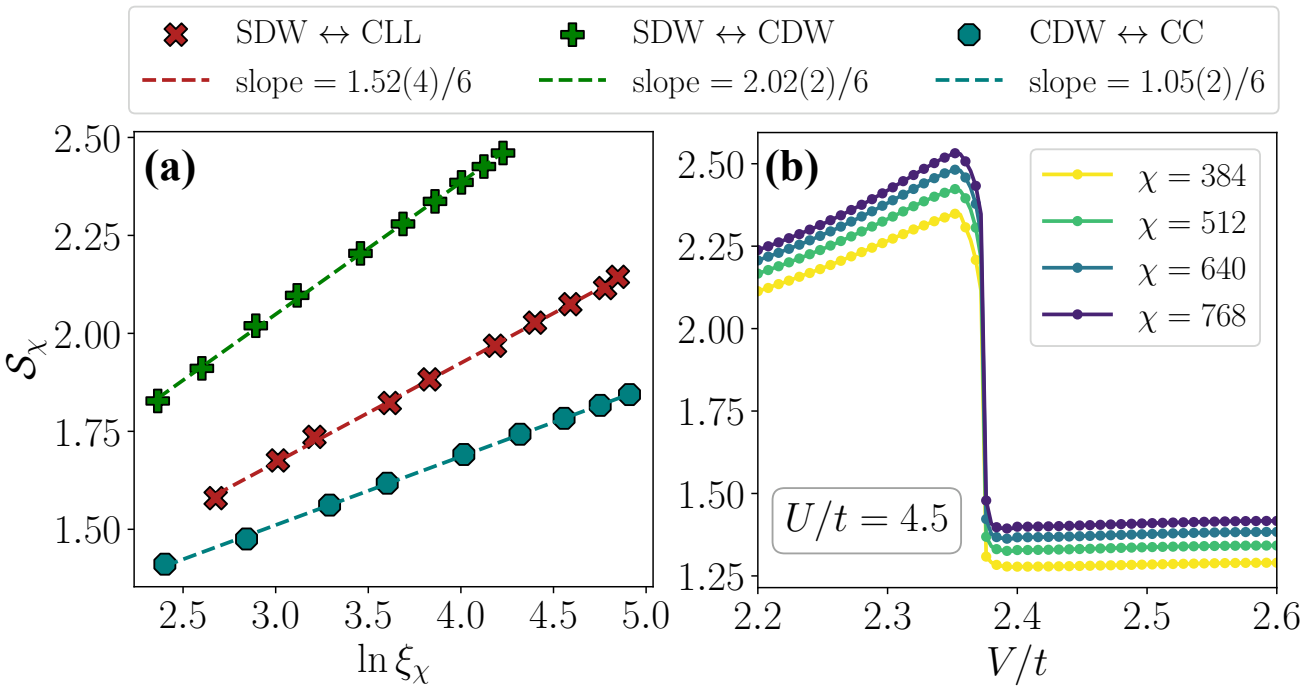

FIG. 3. (Color online.) **(a)** The scaling of the entanglement entropy $\mathcal{S}_\chi$ with the correlation length $\xi_\chi$ according to Eq. (2) for different phase transitions. We have tracked the transitions by fixing $U/t = 1$ (for SDW $\leftrightarrow$ CLL), $U/t = 3.5$ (for SDW $\leftrightarrow$ CDW), $U/t = 4.5$ (for CDW $\leftrightarrow$ CC), and varying $V/t$ (see Ref. [18]). **(b)** The variations of the entanglement entropy $\mathcal{S}_\chi$ across the CDW $\leftrightarrow$ CLL transition for different bond dimensions $\chi$. The sudden jumps in $\mathcal{S}_\chi$ suggest the the transition to be first order.

and $V/t \lesssim 0.34$, we find a polarized holonic (PH) phase.

To identify and characterize the (gapless) phases and phase transitions, we compute the central charges $c$ of each by evaluating how $S_\chi$ scales with $\xi_\chi$ using [37–39]:

$$\mathcal{S}_\chi = \frac{c}{6} \ln \xi_\chi + b', \tag{2}$$

where $b'$ is a non-universal constant. We thereby extract $c = 1$ for the CLL, SDW, and CDW phases, while $c = 2$ for the 2TLL phase. We refer to the supplementary material [21] for both the associated scalings of entanglement entropy and correlation functions characterizing these phases by their leading instabilities.

Moreover, following the method used in Refs. [18, 32], we identify the $c = 3/2$ SDW $\leftrightarrow$ CLL, Gaussian $c = 2$ SDW $\leftrightarrow$ CDW, and the $c = 1$ CDW $\leftrightarrow$ CC phase transitions (see Fig. 3**(a)**). We expect the BKT phase transition between the SDW and the 2TLL phases by analogy with the square case. The sudden discontinuities in $\mathcal{S}_\chi$ seen in Fig. 3**(b)** indicate that the direct CDW $\leftrightarrow$ CLL transition is of the first order type.

As the iDMRG simulations do not converge properly in the PH phase due to frustrations originating from strong repulsive interaction in the triangular geometry, we cannot characterize the PH phase or the PH $\leftrightarrow$ CC and PH $\leftrightarrow$ CDW phase transitions from the iDMRG results. Therefore, to characterize both the PH and the CC phases, we use a semi-classical strong-coupling approach, completed by exact diagonalization when possible.

We note that the 2TLL $\leftrightarrow$ CDW transition is too narrow to be well studied numerically. It is also too deep in the intermediate interaction regime to be approached by perturbative techniques, hence the dotted lines in Fig. 1 (b).

*Strong intra-chain coupling.* To provide analytical predictions for the large $U$ regime, we use a perturbative strong coupling approach where $t/U$ is the perturbation. Semi-classically in the limit $U \to \infty$, both the geometry and the strong intra-chain interaction segment the ladder into polarized domains. Each domain sees one leg of the ladder hosting atoms while the opposite leg is empty. States displaying such domains as in Fig. 4**(a)** and **(b)** generate the effective Hilbert space at strong-coupling. The filled leg of the ladder within a domain may host (quasi-)holes, that we call polarized holons: these holons freely propagate within the domain, and may hop from one domain to the next at no energy cost. The density of such holons $f_h$ and the density of domain walls $f_w$ is fixed by the filling $\nu$, such that

$$f_w + 2f_h = 2 - 4\nu, \tag{3a}$$

$$0 \leq f_w \leq \mathrm{Min}(2\nu, 2 - 4\nu). \tag{3b}$$

When $V/t$ is negligible, configurations that maximize the density of holons and minimize the number of domains are favored energetically as the holons can delocalize both within and beyond a domain. It is the polarized holonic (PH) liquid. Instead, for larger $V$, small but numerous domains (i.e., clusters in that case) are energetically favored. Deprived of the propagating holons, the system is fully gapped. We denote this regime as the cluster crystal. When $V \geqslant 2U$, the strong $U$ coupling approach breaks down.

To derive these results, we first obtain the perturbative Hamiltonian at second order in perturbation in $t^2/U$ with $V = 0$. Restricted to the effective Hilbert space of the strong coupling, the perturbative Hamiltonian is

$$\begin{aligned} H|_{\mathrm{res}} = t \sum_{i=1}^{L} \sum_{\ell} (b^\dagger_{i,\ell} b_{i+1,\ell} + \mathrm{H.c.}) \\ - \frac{2t^2}{U} \sum_{i} (n_{i,+} n_{i+2,-} + n_{i,-} n_{i+1,+}) \,. \end{aligned} \tag{4}$$

In this Hilbert space, the first term of Eq. (4) corresponds to the hopping of one particle *two* sites further along the zigzag if there are immediately at least three consecutive empty sites in the direction of the hopping (see Fig. 4**(a)**). These three consecutive holes constitute the holon. When a holon is immediately followed by other holons (its cohort), each of these extra holons adds two holes along the zigzag. In this formulation, the first term of Eq. (4) simply describes the hopping of free holons. The second term of Eq. (4) is a chemical potential for succession of exactly two empty sites. The polarization of the ladder changes across such a structure: it is a domain wall. A domain wall can also accompany a cohort of consecutive holons, each of them adding two holes along the zigzag. By convention, we consider that the domain wall is always at the left of the cohort.

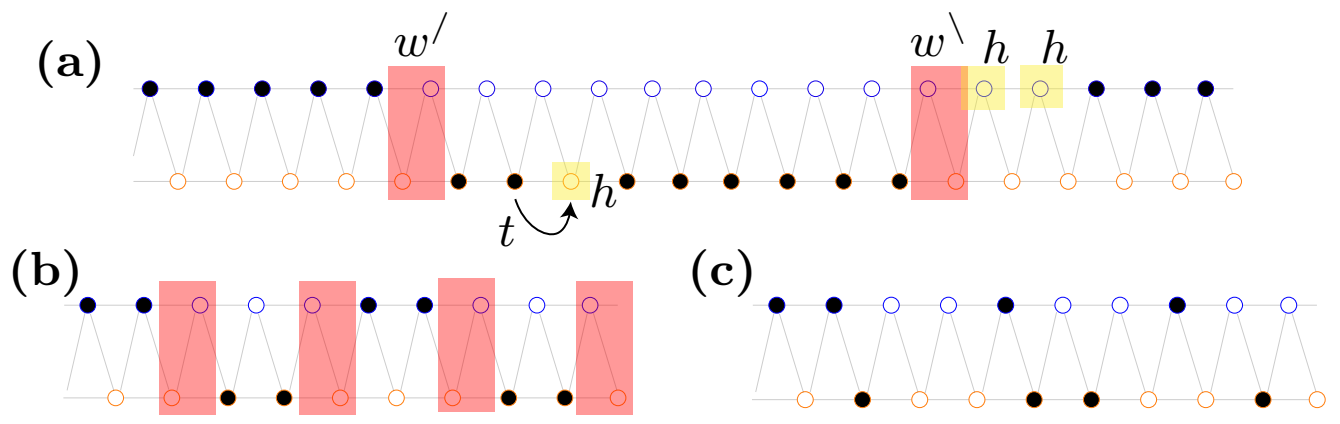

FIG. 4. (Color online.) **(a)** The cartoon picture of the PH phase and example of a semi-classical state of the effective Hilbert space when $U \to \infty$. The ladder is divided into polarized segments (here, only two with periodic boundary conditions). These segments are separated by domain walls (in red). Holes in the segments are holons (in yellow). Holons can belong to the cohort of a domain wall, like for the left domain wall on the picture. Arrow: example of hopping. **(b)** The cartoon picture of the CC phase. **(c)** One semi-classical configuration of the CLL phase.

To rewrite the effective model in terms of holons and domain walls, we require a creation and annihilation operator for both the holons ($h_i^\dagger$ and $h_i$) and the domain walls ($w_i^\dagger$ and $w_i$). It is possible to derive the explicit expression for both the densities of holons ($n_{hi}$) and domain walls ($n_{wi}$) (see [21]). The existence of these densities implies the existence of the associated creation and annihilation operators. Using these operators, we define the basis of the effective Hilbert space such that

$$h_i^\dagger|0_i\rangle = |1_i\rangle, \qquad h_i|1_i\rangle = |0_i\rangle, \tag{5a}$$
$$h_i^\dagger|1_i\rangle = 0, \qquad h_i|0_i\rangle = 0, \tag{5b}$$
$$w_i^\dagger|X_i\rangle = |w_i X_i\rangle, \qquad w_i|w_i X_i\rangle = |X_i\rangle, \tag{5c}$$

with $X = 0$ or $1$, and $\{h_i, h_i^\dagger\} = \{w_i, w_i^\dagger\} = 1$. For simplicity, we assume that the holonic and domain walls operators commute, essentially treating them as decoupled degrees of freedom. Such approximation is justified a posteriori, by comparing our findings to numerics.

Using these operators, we obtain a phenomenological expression for the perturbative Hamiltonian from which we deduce an estimate of the energy levels as a function of the average density of domain walls and holon. The Hamiltonian is

$$\begin{aligned} H|_{\rm res} \sim t \sum_{i=1}^{L} \Big[ & w_i^\dagger h_{i+1}^\dagger w_{i+1} h_i + h_i^\dagger w_{i+1}^\dagger h_{i+1} w_i \\ & + h_i^\dagger h_{i+1} (1 - n_{w,i+1}) + h_{i+1}^\dagger h_i (1 - n_{w,i+2}) \Big] \\ & - \frac{2t^2}{U} \sum_i w_i^\dagger w_i h_i h_i^\dagger . \end{aligned} \tag{6}$$

In Eq. (6), the first four terms describe the hopping of holons depending on the proximity of a wall. When no wall is nearby, the hopping holon is unimpeded. When a holon hops across a wall, the holon and wall exchange places. The last term in Eq. (6) counts the number of walls without cohort. As a consequence, a wall effectively moves only through scattering with a holon. All walls hop one site when a holon propagates along the entire ladder. Due to this difference in scale of the two momenta, we work under the assumption that the two variables are independent to estimate the spectrum of the Hamiltonian. We take $\langle n_{hi} \rangle = f_h L$ and $\langle n_{wi} \rangle = f_w L$. We find

$$\begin{aligned} \langle H|_{\rm res} \rangle / 2tL = \frac{t}{U} f_w (f_h - 1) - \frac{1}{\pi}(1 - f_w) \sin(\pi f_h) \\ - \frac{2}{\pi^2} \sin(\pi f_h) \sin(\pi f_w / 2). \end{aligned} \tag{7}$$

Eq. (7) highlights competition between holons and domain walls. When $U/t > 2\pi/(5\sin(\pi/5)) \sim 2.14$, the phase is the polarized holonic liquid with $c = 1$. Such a small value of $U/t$ is beyond the validity of the strong-coupling approach which explains the difference with the value of $\sim 3.1$ observed in Fig. 2. Third order corrections in $U/t$ and higher allow the transformation of two domain walls of small domains into one holon and vice-versa.

Similarly, we obtain the contribution of the shoulder potential to the estimated energy levels using the holons and domain walls reformulation. We find (see [21])

$$\begin{aligned} & \langle H_{\rm Shoulder} \rangle / VL = \\ & \quad 2 - 4f_h - 4f_w + 2f_h^2 - \frac{1}{\pi^2}\sin^2(\pi f_h) \\ & \quad - \frac{1}{2\pi^2} \sin(\pi f_h) \sin(2\pi f_h) + 4 f_h f_w + \frac{1}{2} f_w^2 . \end{aligned} \tag{8}$$

The full estimate of the energy levels is the sum of Eqs. (7) and (8). Because of Eq. (3a), the maximal domain wall density is twice the maximal holon density, such that Eq. (8) unilaterally favours the domain walls over the holons. A transition thus occurs when the contribution of Eq. (8) dominates over the contribution from Eq. (7). When $U \to \infty$, this transition is predicted at $V/t \sim 0.33$ and observed at $V/t \sim 0.34$ on Fig. 2. Higher order corrections effectively introduce a repulsion between domain walls. When the density of domain walls is maximized, a pattern spontaneously breaking the translation symmetry emerges, and the system goes into the CC phase.

By comparing the energy of semi-classical configuration of the CC pattern as in Fig. 4**(b)** $E_{CC} = 2V$ and of a CLL state minimizing $U$ as in Fig. 4 **(c)** (not included in the restricted Hilbert space) $E_{CLL} = V + 2U$ when $t = 0$, we predict the end of the CC phase at $V = 2U$. Due to the existence of the CDW phase (CDW are not included in the restricted Hilbert space), the transition occurs at smaller $V < 2U$ for finite $U$.

We extract the order parameter $\mathcal{O}_{CC}$ associated with the CC phase using the structure factor:

$$S(q,k) = \frac{1}{L^2} \sum_{j,j'} \sum_{\ell,\ell'} \langle n_{j,\ell} n_{j',\ell'} \rangle e^{-i(q + k(\ell - \ell'))(j - j')} . \tag{9}$$

In the CC phase at $\nu = 2/5$, $S(q,k)$ displays a sharp peak at momenta $q = 2\pi/5, k = 2\pi/5$ (see Fig. 5**(a)**). Sub-tracting the disconnected contribution from $S(q,k)$, we define the order parameter $\mathcal{O}_{CC} = S(2\pi/5, 2\pi/5) - 4/25$ of the CC phase. Figs. 5**(b)** and **(c)** show the variation of the order parameter along the CC $\leftrightarrow$ PH and the CC $\leftrightarrow$ CDW phase transitions respectively. The sharp changes in $\mathcal{O}_{CC}$ across the CC $\leftrightarrow$ PH transition, even for small system-size as $L = 10$, strongly hint that this transition might be first order in nature.

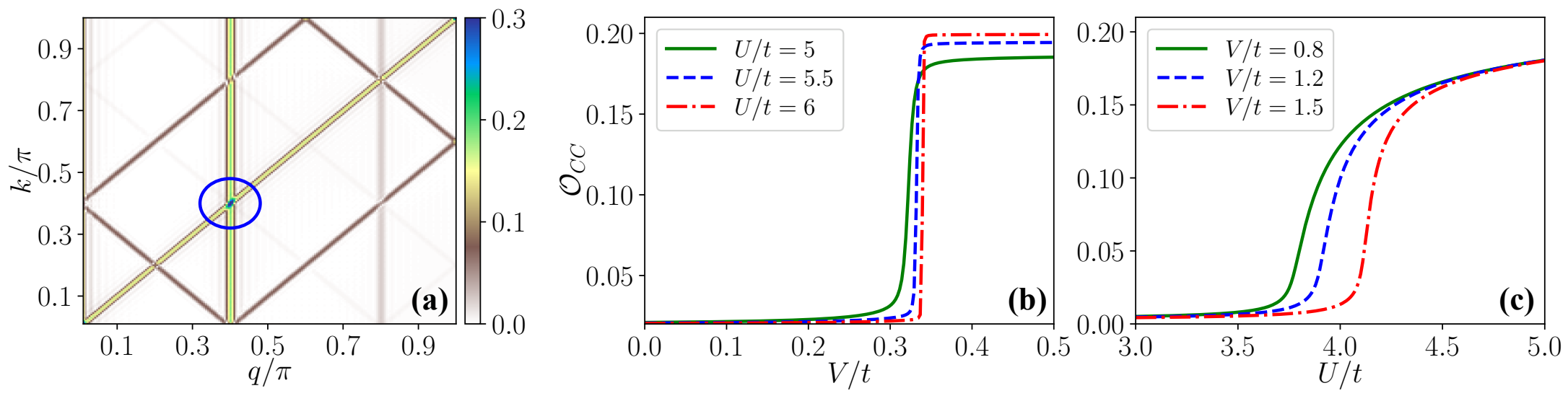

FIG. 5. (Color online.) **(a)** The structure factor $S(q,k)$ (Eq. (9)) in the CC phase ($U/t = 5, V/t = 1.5$) as a function of the momenta $q$ and $k$. $S(q,k)$ attains a sharp peak at $q = 2\pi/5, k = 2\pi/5$ as highlighted by the blue circle. **(b)**-**(c)** The order parameter $\mathcal{O}_{CC}$ (see text) along the the CC $\leftrightarrow$ PH and the CC $\leftrightarrow$ CDW phase transitions respectively. The figures **(a)** and **(c)** have been obtained from iDMRG simulations with bond dimension $\chi = 256$, while **(b)** is obtained from ED calculations for a system with 10 sites.

*Conclusions.* We have determined the phase diagram of a triangular ladder of hardcore bosons at $\nu = 2/5$ filling with next-to nearest neighbor longitudinal interaction and no transversal hopping. The model is inspired by experimental setups using Rydberg atoms loaded in an optical lattice [17], where most of the regimes we propose are in principle accessible. Similar physics shall be accessible in dressed regimes in tweezers [40]. The tunneling between the chains can be crucially suppressed by leveraging on the comparatively large quadratic Zeeman shifts from chain to chain. Alternatively, one can use a high potential barrier between the chains.

We have seen that the extension of the range of the inter-leg interaction immediately redesigns the vicinity of the supersymmetric point with respect to the square ladder case. In particular, a single phase transition line, likely to be a supersymmetric one, separates the weak and strong intra-chain interacting regimes at weak inter-chain interaction. The extra range only deforms the neighboring SDW and spin-locked CLL instead. Such results illustrate the difficulty in controlling the coupling between two supersymmetric field theories like in the square case as power-law tails always exist in realistic systems. Instead, the CLL phenomenology is robust to the inter-leg interaction range extension.

The strong coupling limit is also richer in the triangular geometry: a CDW appears at intermediate $U/t$, and the PH and gapped translation symmetry breaking CC emerge at large $U/t$, compared to the square case. The gapless PH is qualitatively unchanged by varying the range of the interaction or the filling. Instead, the CC disappears as soon as the filling (2/5 here) is changed or the interaction range is further extended akin to gapped phases in half-filled Hubbard models. We thus expect a plethora of other CC patterns for other ranges of $V$ and matching commensurate filling, separated at incommensurate filling by both standard and cluster Luttinger liquid phases. The phenomenology of the extended Hubbard model at any filling and at strong coupling is thus more complex than the sole clusterization of the degrees of freedom but involves instead the coexistence of phases with both clustered and Luttinger bosonic excitations separated by uncommon (*e.g.,* supersymmetric) phase transitions.

*Acknowledgments.* We acknowledge useful discussions with A. Angelone, M. Fabrizio, G. Giudici, G. Japaridze and P. Lecheminant. The work of M. D., P. F. and M. T. is partly supported by the ERC under grant number 758329 (AGEnTh), by the MIUR Programme FARE (MEPH), and by the European Union's Horizon 2020 research and innovation programme under grant agreement No 817482 (Pasquans). M. T. thanks the Simons Foundation for supporting his Ph.D. studies through Award 284558FY19 to the ICTP. P.F. and part of this work were supported by the Swiss National Science Foundation, and NCCR QSIT (Grant number 51NF40-185902). The work by A.N. was partly supported by the Shota Rustaveli National Science Foundation of Georgia, SRNSF, Grant No. FR-19-11872. The iDMRG simulations have been performed using the TeNPy library [41]. The code used to obtain the data presented in the paper is directly adapted from Ref. [42].

* pierre.fromholz@unibas.ch
† mtsitsis@ictp.it
‡ tchanda@ictp.it

# Supplemental Material: Phase diagram of Rydberg-dressed atoms on two-leg triangular ladders

Pierre Fromholz,[1] Mikheil Tsitsishvili,[2,3] Matteo Votto,[2,3] Marcello Dalmonte,[2,3] Alexander Nersesyan,[2,4,5] and Titas Chanda[2]

[1] *Department of Physics, University of Basel, Klingelbergstrasse 82, CH-4056 Basel, Switzerland*
[2] *The Abdus Salam International Centre for Theoretical Physics (ICTP), Strada Costiera 11, 34151 Trieste, Italy*
[3] *International School for Advanced Studies (SISSA), via Bonomea 265, 34136 Trieste, Italy*
[4] *The Andronikashvili Institute of Physics, 0177 Tbilisi, Georgia*
[5] *Ilia State University, 0162 Tbilisi, Georgia*

## I. EFFECT OF THE TRIANGULAR INTERACTION ON THE SUPERSYMMETRIC CRITICAL POINT

Using the infinite density-matrix renormalization group (iDMRG) [1–8] techniques mentioned in the main text, we study the phase diagram in the vicinity of the supersymmetric (SUSY) $c = 3$ critical point at $V/t \sim 5.7$ and $U = U_1 = 0$ as $U_1$ is tuned from 0 (square geometry) to $U$. We observe that as soon as $U_1 > 0$ ($U_1 = U/10$ in our simulation), the Charge Density Wave (CDW) phase existing in the square geometry [9] vanishes (see Fig. 1). Along with the phase, the $c = 2$ Gaussian phase transition existing between the Spin Density Wave (SDW) and the CDW vanishes as well. We interpret this result as the non-zero triangular term being relevant for these parameters, preventing the phase transition and deforming the SDW phase into the SDW($\alpha$) (see Sec. II).

While no Gaussian transition appears in our simulation, a more detailed study in its vicinity shows that a remnant of it still exists (see Fig. 2). Such observation is compatible with a spin gap reaching a local *finite* minimum corresponding to both a correlation length $\xi_\chi$ and entanglement entropy $\mathcal{S}_\chi$ reaching a local maximum. The absence of a sharp peak (present for the transition in the square case, see Ref. [9]) confirms the absence of the transition.

The $c = 3/2$ phase transition between the spin-locked Cluster Luttinger Liquid (CLL) and the SDW phases remains. The phenomenological understanding of this phase transition as the sign inversion of the mass of an effective Majorana field existing at the SUSY point for the square geometry [9] holds as well for the triangular geometry.

## II. WEAK COUPLING APPROACH

Using the Abelian bosonization [10, 11] method we study the phase diagram of the triangular ladder model in the weak coupling regime. Specifically, after reformulating the description of the system in terms of fermions using a Jordan-Wigner transformation, we take the continuous limit of the discrete model, and linearize it around the Fermi points. To the obtained chiral Dirac fermions, we then apply the standard Abelian bosonization relations to finally derive the sine-Gordon model of two compact bosons. Using a conformal renormalization scheme at first loop, we deduce the asymptotes of the renormalization group equations depending on the initial conditions. We therefore obtain the outline of the phase diagram at weak coupling. By computing correlation functions associated to each asymptote, we identify each phase of this diagram. Using this approach, we predict the gapless 2TLL and the $\mathrm{SDW}^z_\alpha$ phases separated by a BKT phase transition.

We perform a Jordan-Wigner transformation to use fermionic degrees of freedom and notions associated with Hubbard-like models:

$$b_{i,\ell} = \left(\delta_{\ell,+} e^{i\pi \sum_{j=1}^{i-1} n_{j,+}} c_{i,+} + \delta_{\ell,-} e^{i\pi \sum_{k=1}^{N} n_{k,+}} e^{i\pi \sum_{j=1}^{i-1} n_{j,-}} c_{i,-}\right), \tag{S.1}$$

where $b_{i,\ell}$ is the hard-core bosonic annihilation operator on site $i$ and leg $\ell = \pm$. The fermionic annihilation operators $c_{i,\ell}$ obey the standard anti-commutation relations. The particle number operators are $n_{i,\ell} = c^\dagger_{i,\ell} c_{i,\ell} = b^\dagger_{i,\ell} b_{i,\ell}$. After the transformation, the triangular ladder Hamiltonian Eq. (1) is unchanged:

$$H = H_0 + H_U + H_{U_1} + H_V, \tag{S.2}$$

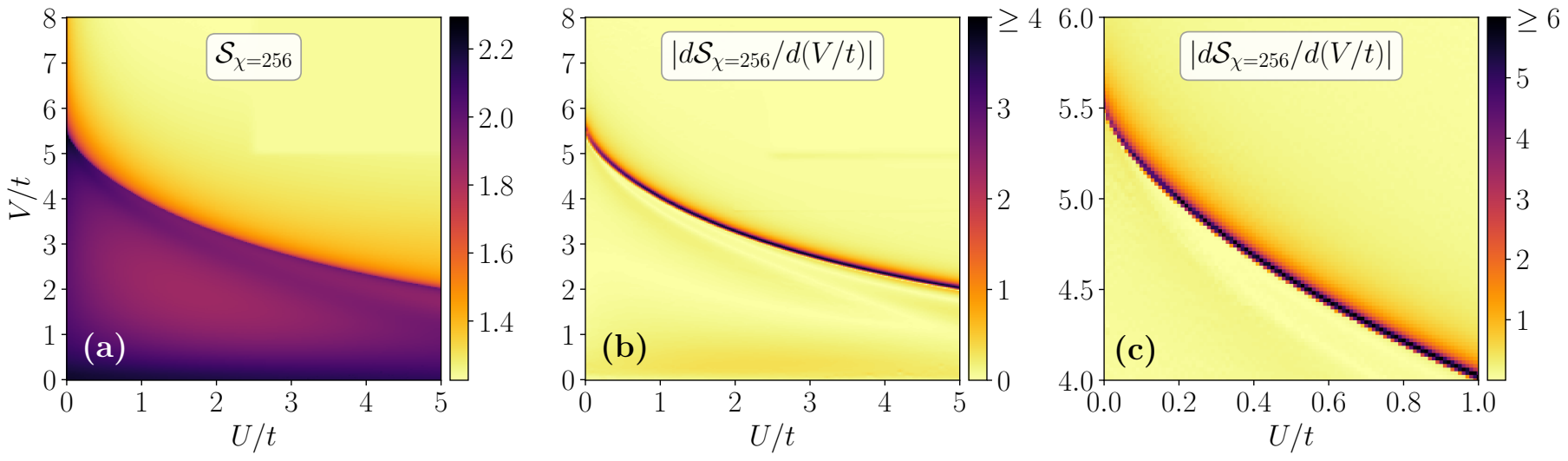

FIG. 1. (Color online.) **(a)** The entanglement entropy $\mathcal{S}$ and **(b)**-**(c)** its derivative with respect to the system parameter $V/t$ in the $(U/t, V/t)$-plane for $U_1 = U/10$. The $c = 3/2$ phase transition is clearly visible while only an imprint of the Gaussian transition, existing for $U_1 = 0$, is present. The quantities are computed by the iDMRG algorithm with bond dimension $\chi = 256$.

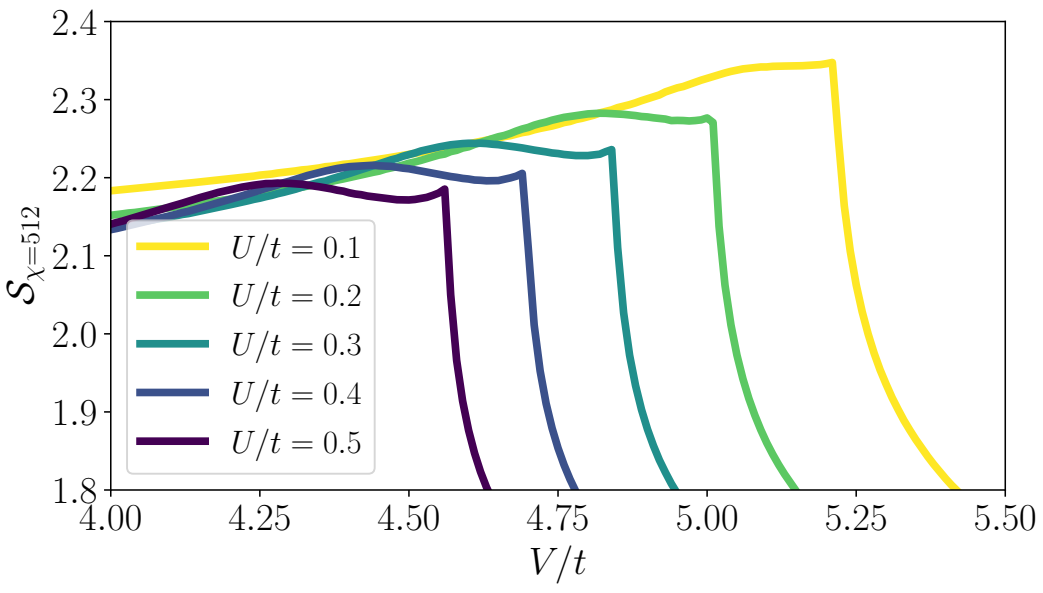

FIG. 2. (Color online.) The variations of the entanglement entropy $\mathcal{S}$ with respect to $V/t$ for various $U/t$ and $U_1 = U/10$. The quantities are computed by the iDMRG algorithm with bond dimension $\chi = 512$. The SDW to CLL phase transition is sharp, whereas the remnant of the Gaussian phase transition is not (compare Fig. 5 in Ref. [9]).

with

$$H_0 = -t \sum_{\ell=\pm} \sum_i \left( c^\dagger_{i,\ell} c_{i+1,\ell} + \text{h.c.} \right), \tag{S.3a}$$

$$H_U = U \sum_i n_{i,+} n_{i,-}, \tag{S.3b}$$

$$H_{U_1} = U_1 \sum_i n_{i,+} n_{i+1,-}, \tag{S.3c}$$

$$H_V = V \sum_{\ell=\pm} \sum_{r=1}^{r_c} n_{i,\ell} n_{i+r,\ell}, \tag{S.3d}$$

where we take $r_C = 2$ unless otherwise specified.

We then linearize the model Eq. (S.2) around the Fermi points (at momentum $\pm k_F$) in the weak coupling regime after taking the continuum limit by introducing the right- and left-moving Dirac fermion fields

$$c_{i,\ell} \sim \psi_{R,\ell}(x = i) + \psi_{L,\ell}(x = i), \tag{S.4a}$$

$$\psi_{r,\ell} = \frac{1}{\sqrt{2\pi a_0}} U_{r,\ell} e^{irk_F x} e^{i\sqrt{\frac{\pi}{2}}(r\phi_c - \theta_c + \ell(r\phi_s - \theta_s))}. \tag{S.4b}$$

$\psi_{r,\ell}$ is the associated right ($r = R$) and left ($r = L$) moving fermionic field in the continuous limit close to the Fermi points, $a_0 = 1$ is the lattice spacing, $U_{r,\ell}$ are the Klein factors such that $\{U_{r,\ell}, U_{r',\ell'}\} = 2\delta_{r,r'}\delta_{\ell,\ell'}$, and $k_F = \pi\nu/a_0$ is the Fermi momentum with $\nu = 2/5$ being the filling factor for each chain. The bosonic fields $\phi_c$, $\phi_s$, $\theta_c$, and $\theta_s$ are the charge and spin fields and their respective dual fields. After bosonization, each term in Eq.(S.3a-S.3d) splits into the spin and charge sector as

$$H_0 \sim \frac{v_F}{2}\int dx\left((\partial_x\theta_c)^2+(\partial_x\phi_c)^2\right)+\frac{v_F}{2}\int dx\left((\partial_x\theta_s)^2+(\partial_x\phi_s)^2\right), \tag{S.5a}$$

$$H_U \sim \frac{U}{2\pi}\int dx(\partial_x\phi_c)^2-\frac{U}{2\pi}\int dx(\partial_x\phi_s)^2+\frac{U}{2\pi^2}\int dx\cos\left(\sqrt{8\pi}\phi_s\right), \tag{S.5b}$$

$$H_{U_1} \sim \frac{U_1}{2\pi}\int dx(\partial_x\phi_c)^2-\frac{U_1}{2\pi}\int dx(\partial_x\phi_s)^2+\frac{U_1(1+\cos(2\pi\nu))}{2\pi^2}\int dx\cos\left(\sqrt{8\pi}\phi_s\right) + \frac{U_1\sin(2\pi\nu)}{2\pi^2}\int dx\sin\left(\sqrt{8\pi}\phi_s\right), \tag{S.5c}$$

$$H_V \sim \frac{V(2-\cos(2\pi\nu)-\cos(4\pi\nu))}{\pi}\int dx(\partial_x\phi_c)^2+\frac{V(2-\cos(2\pi\nu)-\cos(4\pi\nu))}{\pi}\int dx(\partial_x\phi_s)^2, \tag{S.5d}$$

with $v_F=2t\sin(\pi\nu)$. The full Hamiltonian displays the spin-charge separation:

$$H = H_c+H_s, \tag{S.6a}$$

$$H_c = \frac{v_c}{2}\int dx\left(K_c(\partial_x\theta_c)^2+\frac{1}{K_c}(\partial_x\phi_c)^2\right), \tag{S.6b}$$

$$H_s = \frac{v_s}{2}\int dx\left(K_s(\partial_x\theta_s)^2+\frac{1}{K_s}(\partial_x\phi_s)^2\right)-\frac{g_0}{4\pi^2}\int dx\cos\left(\sqrt{8\pi}\phi_s-\alpha\right), \tag{S.6c}$$

where $H_c$ is the bosonized Hamiltonian of the charge sector, and $H_s$ is the sine-Gordon Hamiltonian of the spin sector. We find

$$K_s=\frac{1}{\sqrt{1+\frac{g_\parallel}{2\pi v_F}}},\qquad v_s=v_F\sqrt{1+\frac{g_\parallel}{2\pi v_F}},\qquad v_sK_s=v_F, \tag{S.7a}$$

$$K_c=\frac{1}{\sqrt{1+\frac{2g}{\pi v_F}}},\qquad v_c=v_F\sqrt{1+\frac{2g}{\pi v_F}},\qquad v_cK_c=v_F, \tag{S.7b}$$

and

$$g \equiv \frac{U+U'}{2}+V(2-\cos(2\pi\nu)-\cos(4\pi\nu)), \tag{S.8a}$$

$$g_\parallel \equiv -2U-2U_1+4V(2-\cos(2\pi\nu)-\cos(4\pi\nu)), \tag{S.8b}$$

$$g_0 = -2\sqrt{(U+U_1)(U+U_1+2U_1\cos(2\pi\nu))+U_1^2}, \tag{S.8c}$$

$$\alpha = \arctan\left(\frac{U_1\sin(2\pi\nu)}{U+U_1(1+\cos(2\pi\nu))}\right)\ \text{modulo } 2\pi. \tag{S.8d}$$

Based on the zeroth loop approach, the cosine perturbation in Eq. (S.6c) is irrelevant as soon as $K_s>1$ (i.e. $g_\parallel<0$). In this regime the spin gap nonetheless opens because of the first-loop corrections to the renormalization group equations [10]

$$\frac{dg_\parallel}{dl}=\frac{1}{2\pi}g_0^2,\qquad \frac{dg_0}{dl}=\frac{1}{2\pi}g_0g_\parallel,\qquad \frac{d\alpha}{dl}=0, \tag{S.9}$$

which predict a BKT phase transition from gapless 2LL to SDW$^z(\alpha)$ with a spin-gap, when $g_\parallel>-|g_0|$. In other words, the zero-th loop predicts the 2TLL $\leftrightarrow$ SDW transition at $V=5U/2$, whereas the first loop renormalization predicts the transition at $V=\frac{4-2\sqrt{4-\sqrt{5}}}{10}U$. Such a result means that the characterization of each phase should be sharp above and below both predicted transition line, but more arduous in between as the partial gap is weaker. Such a phenomena explains why the detection of the BKT phase transition is more difficult in numerical simulations. We draw the $V=5U/2$ line on the phase diagram as it is below this line that a simulation likely mistakenly confuses the SDW for the 2TLL.

In the SDW$^z(\alpha)$ phase the spin field is fixed at $\phi_s(\alpha)=\frac{\alpha}{\sqrt{8\pi}}+\sqrt{\frac{\pi}{8}}$ modulo $\left(\sqrt{\frac{\pi}{8}}\right)$ by the spin gap. The corresponding order parameter is

$$\mathcal{O}_\alpha = \sin\left(\sqrt{2\pi}\phi_s-\alpha/2\right), \tag{S.10a}$$

$$= \mathcal{O}_{\text{triplet}_0}\cos(\alpha/2)-\mathcal{O}_{\text{singlet}}\sin(\alpha/2), \tag{S.10b}$$

with $\alpha \in [0; 2\pi/5]$ as $U_1/U \in [0; \infty)$ and where $\mathcal{O}_{\text{triplet}_0}$ is the "triplet 0" and $\mathcal{O}_{\text{singlet}}$ is the singlet order parameter [12]. When $\alpha = 0$ ($U_1 = 0$), this order parameter corresponds to the contribution of the spin sector to the $2k_F$ spin-density-wave operator

$$\begin{aligned}\mathcal{O}^z_{\text{SDW}} &= S^z(x) - \frac{1}{\sqrt{2\pi}}\partial_x\phi_s(x)\\ &= \cos\Big(\sqrt{2\pi}\phi_c + 2k_F x\Big)\sin\Big(\sqrt{2\pi}\phi_s\Big),\end{aligned} \tag{S.11}$$

where $S^z(x)$ the continuous limit of the spin operator

$$\frac{1}{2}\sum_{\ell,\ell'=\pm} c^\dagger_{x,\ell}\sigma^z_{\ell\ell'}c_{x,\ell'} \sim S^z(x), \tag{S.12}$$

with $\sigma^z_{\ell\ell'}$ the Pauli matrix along the $z$-direction. The associated correlation function in $\text{SDW}^z(\alpha)$ phase is

$$\langle\mathcal{O}^z_{\text{SDW}}(x)\mathcal{O}^z_{\text{SDW}}(y)\rangle \sim \sin^2\left(\sqrt{2\pi}\phi_s(\alpha)\right)\cos(2k_F|x-y|)|x-y|^{-K_c}, \tag{S.13}$$

which is also the correlator with the longest range, such that the $\text{SDW}^z(\alpha)$ is the dominant instability in the system.

In 2TLL phase, both $2k_F$ momentum CDWs and SDWs instabilities are in competition. The order parameter associated with CDW is defined as

$$\mathcal{O}_{\text{CDW}} = \rho(x) - \sqrt{\frac{2}{\pi}}\partial_x\phi_c(x), \tag{S.14}$$

where $\rho(x)$ is the continuous limit of the density operator

$$\sum_{\ell=\pm} c^\dagger_{x,\ell}c_{x,\ell} \sim \rho(x). \tag{S.15}$$

The correlation functions associated with both CDW and $\text{SDW}^z$ are

$$\langle\mathcal{O}_{\text{CDW}(x)}\mathcal{O}_{\text{CDW}}(y)\rangle \sim \cos(2k_F|x-y|)|x-y|^{-(K_c+K^*_s)}, \tag{S.16a}$$

$$\langle\mathcal{O}^z_{\text{CDW}}(x)\mathcal{O}^z_{\text{CDW}}(y)\rangle \sim \cos(2k_F|x-y|)|x-y|^{-(K_c+K^*_s)}, \tag{S.16b}$$

with $K^*_s = 1$, the renormalized Luttinger parameter as the SU(2) symmetry emerges asymptotically.

## III. CLUSTER BOSONIZATION APPROACH

The low-energy effective theory of a single chain ($U = U_1 = 0$) at strong coupling $V \gg t$ is effectively described by cluster bosonic fields, *i.e.* density fluctuations of perfect cluster configuration of particles. To derive this result, we use the Cluster bosonization approach. In the continuum limit, the bosonic field operators are expressed as [9, 13]

$$\psi_\ell(x) = e^{-\frac{i\sqrt{\pi}}{\sigma}\theta'_\ell(x)}\sqrt{\nu - \frac{1}{\sqrt{\pi}\sigma}\nabla\phi'_\ell(x)}\sum_{q=-\infty}^{+\infty}\alpha_{q,\ell}e^{-i2q(\pi x\nu\sigma - \sqrt{\pi}\phi'_\ell(x))} \tag{S.17}$$

and the particle density operator as

$$\rho_\ell(x) = \left(\nu - \frac{1}{\sqrt{\pi}\sigma}\nabla\phi'_\ell(x)\right)\sum_{q=-\infty}^{+\infty}A_{q,\ell}e^{-i2q(\pi x\sigma\nu - \sqrt{\pi}\phi'_\ell(x))}. \tag{S.18}$$

$\phi'_\ell(x)$ is the fluctuation above a perfect cluster configuration on chain $\ell$. $\theta'_\ell(x)$ is defined as the conjuguate variable of $\nabla\phi'_\ell(x)$ such that

$$\left[\theta'_\ell(y), \frac{1}{\pi}\nabla\phi'_{\ell'}(x)\right] = i\delta_{\ell,\ell'}\delta(x-y). \tag{S.19}$$

$A_{q,\ell} = A^*_{-q,\ell}$ and $\alpha_{q,\ell}$ are the non-universal constants that strongly depend on the microscopic details of the system. $\sigma = M/N$ with $M$ and $N$ being the number of clusters and particles respectively.

In the absence of the inter-chain interactions $U$ and $U_1$, we have two decoupled chains with the soft-shoulder interaction $V$ acting within each chain. Using the cluster bosonization, the Hamiltonian of each chain is mapped to the Hamiltonian of free massless cluster bosons [13, 14]

$$H_{V,t} \sim \sum_{\ell=\pm} \frac{v}{2} \int dx \left( (\nabla\varphi'_\ell)^2/K + K(\nabla\theta'_\ell)^2 \right), \tag{S.20}$$

where $K$ is the cluster-Luttinger parameter. Since the top and bottom chains are identical, we have $K_+ = K_- = K$ and $v_+ = v_- = v$. By dropping the constants and only keeping the most relevant terms, the mapping for the $U$ interaction is [9]

$$\begin{aligned} H_U &\sim U \int dx \rho_+(x)\rho_-(x), \\ &\approx g_U^{(0)} \int dx \nabla\varphi'_+(x)\nabla\varphi'_-(x) + g_U^{(1)} \int dx \cos\Big(\sqrt{4\pi}[\varphi'_+(x) - \varphi'_-(x)]\Big), \end{aligned} \tag{S.21}$$

with $g_U^{(0)} = \frac{U|A_0|^2}{\pi\sigma^2}$ and $g_U^{(1)} = 2\nu^2 U|A_1|^2$. Similar calculations for $U_1$ interaction gives

$$\begin{aligned} H_{U_1} &\sim U_1 \int dx \rho_+(x)\rho_-(x+a_0), \\ &\approx g_{U_1}^{(0)} \int dx \nabla\varphi'_+(x)\nabla\varphi'_-(x) + g_{U_1}^{(1)} \int dx \cos\Big(\sqrt{4\pi}[\varphi'_+(x) - \varphi'_-(x)]\Big) \\ &\quad - g_{U_1}^{(2)} \int dx \sin\Big(\sqrt{4\pi}[\varphi'_+(x) - \varphi'_-(x)]\Big), \end{aligned} \tag{S.22}$$

with

$$g_{U_1}^{(0)} = \frac{U_1|A_0|^2}{\pi\sigma^2}, \tag{S.23a}$$

$$g_{U_1}^{(1)} = 2\nu^2 U_1 |A_1|^2 \cos(2\pi n\sigma a_0), \tag{S.23b}$$

$$g_{U_1}^{(2)} = 2\nu^2 U_1 |A_1|^2 \sin(2\pi n\sigma a_0). \tag{S.23c}$$

Using the spin and charge cluster bosonic fields

$$\varphi'_\ell = \frac{\varphi'_c + \ell\varphi'_s}{\sqrt{2}}, \qquad \theta'_\ell = \frac{\theta'_c + \ell\theta'_s}{\sqrt{2}}, \tag{S.24}$$

with $\ell = \pm$, the total hamiltonian for the triangular ladder $H = H_{V,t} + H_U + H_{U_1}$ is

$$H = \frac{v_c}{2} \int dx \left( K_c(\nabla\theta'_c)^2 + \frac{1}{K_c}(\nabla\varphi'_c)^2 \right) + \frac{v_s}{2} \int dx \left( K_s(\nabla\theta'_s)^2 + \frac{1}{K_s}(\nabla\varphi'_s)^2 \right) + g \int dx \cos\Big(\sqrt{8\pi}\varphi'_s - \alpha\Big), \tag{S.25}$$

with

$$g = \sqrt{\left(g_U^{(1)} + g_{U_1}^{(1)}\right)^2 + \left(g_{U_1}^{(2)}\right)^2}, \qquad \alpha = \arctan\left(\frac{-g_{U_1}^{(2)}}{g_U^{(1)} + g_{U_1}^{(1)}}\right), \tag{S.26a}$$

$$K_c = \frac{K}{\sqrt{1 + \frac{\left(g_U^{(1)} + g_{U_1}^{(1)}\right)K}{v}}}, \qquad v_c = v\sqrt{1 + \frac{\left(g_U^{(1)} + g_{U_1}^{(1)}\right)K}{v}}, \tag{S.26b}$$

$$K_s = \frac{K}{\sqrt{1 - \frac{\left(g_U^{(1)} + g_{U_1}^{(1)}\right)K}{v}}}, \qquad v_s = v\sqrt{1 - \frac{\left(g_U^{(1)} + g_{U_1}^{(1)}\right)K}{v}}. \tag{S.26c}$$

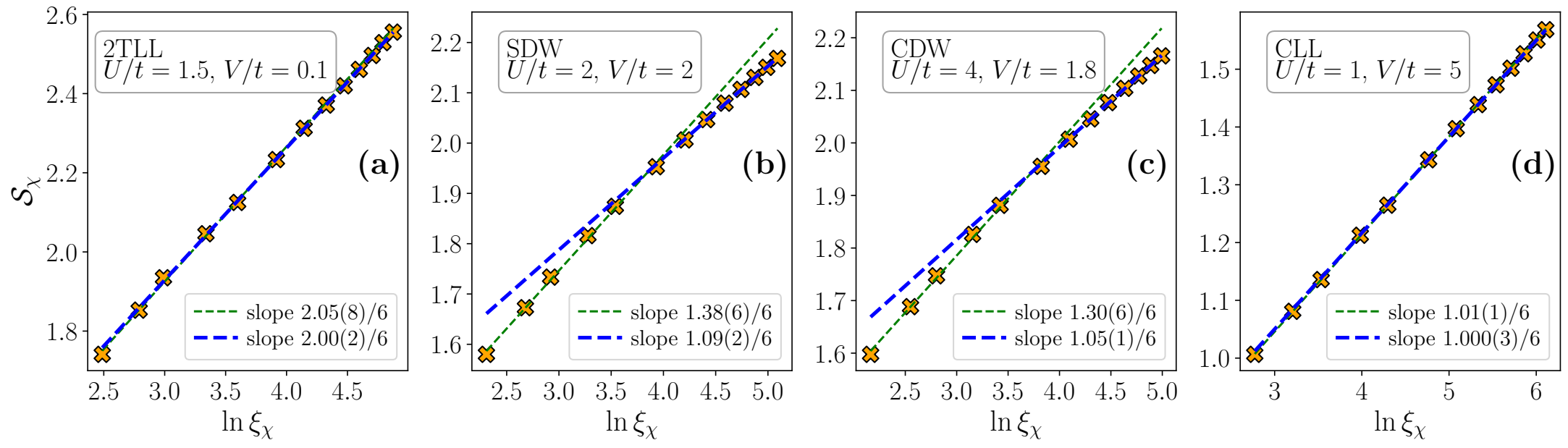

FIG. 3. (Color online.) The behavior of the entanglement entropy $\mathcal{S}_\chi$ with respect to the correlation length $\xi_\chi$ for the bond dimensions $\chi \in [64, 1280]$ in different phases, namely **(a)** 2TLL, **(b)** SDW, **(c)** CDW, and **(d)** CLL. The thin green lines correspond to the fits (see Eq. (S.27)) in the range $\chi \in [64, 384]$, while the thick blue lines correspond to the fits for $\chi \in [512, 1280]$.

As the effective cluster Hamiltonian Eq. (S.25) is formally similar to the bosonized Hamiltonian Eq. (S.6), we conclude that two phases are possible ($g > 0$). If the spin mass is irrelevant, then one possible phase is a CLL with $c = 2$. If instead the mass is relevant, then the second possible phase is a 'spin-locked' CLL with $c = 1$. The microscopic details of the coarse-graining realized with the Cluster bosonization are hidden within the $A_{\ell,q}$ and $\alpha_{\ell,q}$ coefficients, preventing their estimations. In consequence, the relevancy of the mass is only accessible by the numerical simulations. The simulations we perform show that $c = 1$ and therefore that the mass is relevant.

## IV. NUMERICAL RESULTS ON THE PHASES AT WEAK AND INTERMEDIATE INTER-CHAIN COUPLING STRENGTHS

By the means of iDMRG simulations, we now numerically analyze different phases that appear at weak and intermediate inter-chain coupling ($U/t$) strengths, i.e., the 2TLL, the SDW, the CDW, and the CLL phases. To this end, we first extract the central charge $c$ for these phases by fitting the bond dimension $\chi$-dependent entanglement entropy $\mathcal{S}_\chi$ as a function of the $\chi$-dependent correlation length $\xi_\chi$ using the following relation[15–17]:

$$\mathcal{S}_\chi = \frac{c}{6} \ln \xi_\chi + b', \tag{S.27}$$

where $b'$ is a non-universal constant. In Fig. 3, we plot the scaling of $\mathcal{S}_\chi$ with respect to $\xi_\chi$ in these phases as we increase the bond dimension from $\chi = 64$ to 1280. Clearly, all these phases, except the 2TLL, have central charge $c = 1$ indicating gapped spin-sectors while the charge-sectors remain gapless as in the case for square geometry [9]. In the 2TLL phase, $c = 2$ signifies that the both sectors are gapless.

We characterize each phases with $c = 1$ phases by analyzing correlation functions of the form

$$\mathcal{C}_\mathcal{O}(R) = \langle \mathcal{O}_j \mathcal{O}_{j+R} \rangle - \langle \mathcal{O}_j \rangle \langle \mathcal{O}_{j+R} \rangle, \tag{S.28}$$

where $\mathcal{O}_j$ is a local operator. When the distance $R$ is greater than the correlation length $\xi_\chi$ all correlations trivially decay exponentially, restricting the validity of the computed correlation function to $R \approx \xi_\chi$ at maximum.

Instead of considering the spin correlation function $\mathcal{C}_{S^z}$ (where $S^z_j = (n_{j,+} - n_{j,-})/2$) and the charge correlation function $\mathcal{C}_n$ (where $n_j = (n_{j,+} + n_{j,-})$), we find that the bond-spin correlation $\mathcal{C}_{B^S}$ and the bond-charge correlation $\mathcal{C}_{B^C}$ are more suitable to characterize the SDW and the CDW phases. Within each of the pairs ($\mathcal{C}_{S^z}$, $\mathcal{C}_{B^S}$) and ($\mathcal{C}_n$, $\mathcal{C}_{B^C}$), the correlation are largely indistinguishable in theory when the charge sector is not gapped [12]. The bond operators are defined as

$$B^S_j = \left( (b^\dagger_{j,+} b_{j+1,+} + \text{H.c.}) - (b^\dagger_{j,-} b_{j+1,-} + \text{H.c.}) \right)/2, \tag{S.29a}$$

$$B^C_j = \left( b^\dagger_{j,+} b_{j+1,+} + b^\dagger_{j,-} b_{j+1,-} + \text{H.c.} \right). \tag{S.29b}$$

Figs. 4**(a)**-**(b)** show these bond correlation functions in the SDW and the CDW phases for $R \leq 200 < \xi_{\chi=2048}$. Clearly, in the SDW phase, the bond-spin correlation function $\mathcal{C}_{B^S}$ follows power-law and the bond-charge correlation

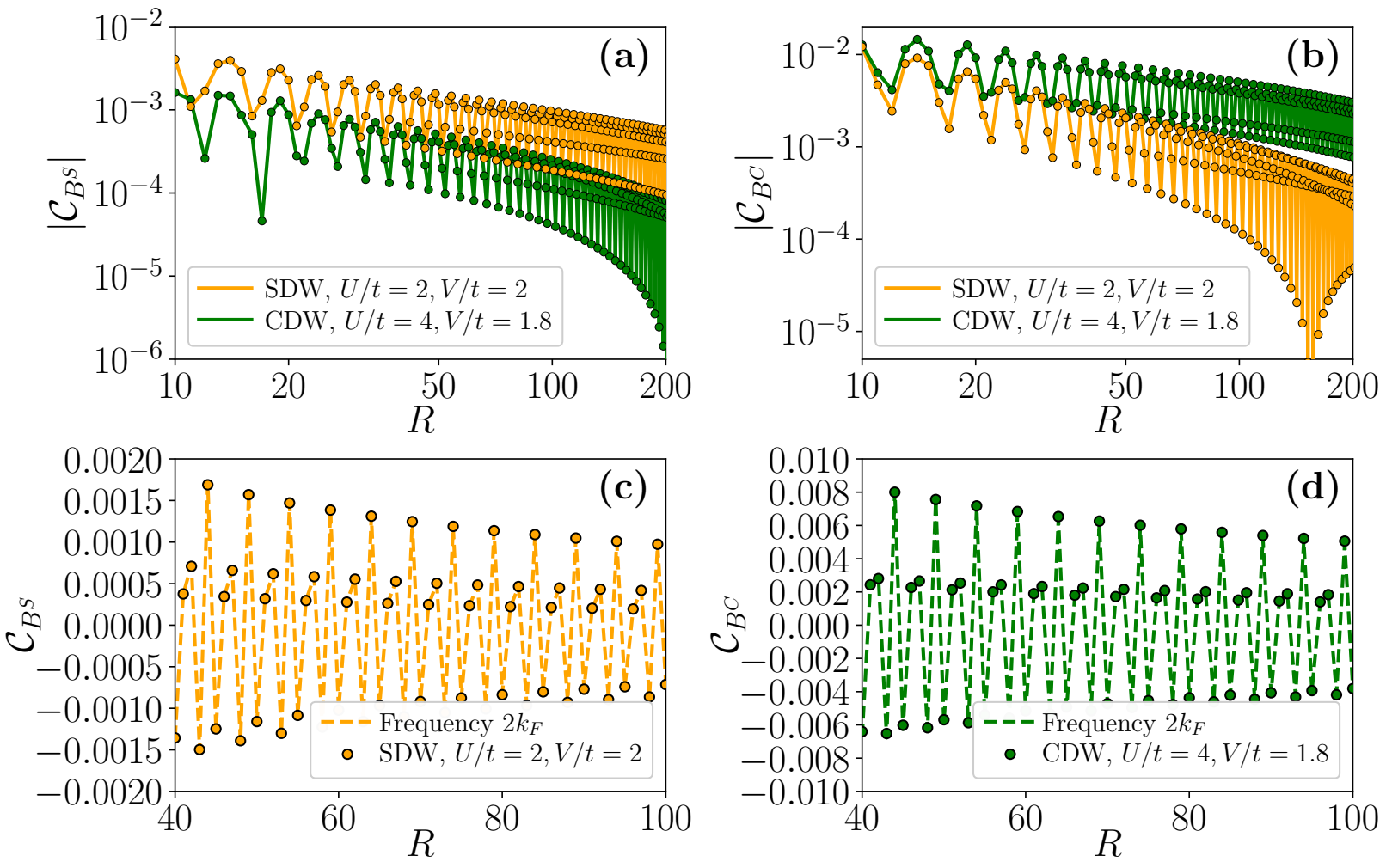

FIG. 4. (Color online.) The behaviors of **(a)** the bond-spin correlation function $\mathcal{C}_{B^S}(R) = \langle B_j^S B_{j+R}^S \rangle - \langle B_j^S \rangle \langle B_{j+R}^S \rangle$ and **(b)** the bond-charge correlation function $\mathcal{C}_{B^C}(R) = \langle B_j^C B_{j+R}^C \rangle - \langle B_j^C \rangle \langle B_{j+R}^C \rangle$ as functions of the distance $R$ in the SDW and CDW phases. Here both axes are in the logarithmic scale. **(c)-(d)** The power-law fits (according to Eq. (S.30)) of the bond-spin and bond-charge correlations respectively in the SDW and CDW phases.

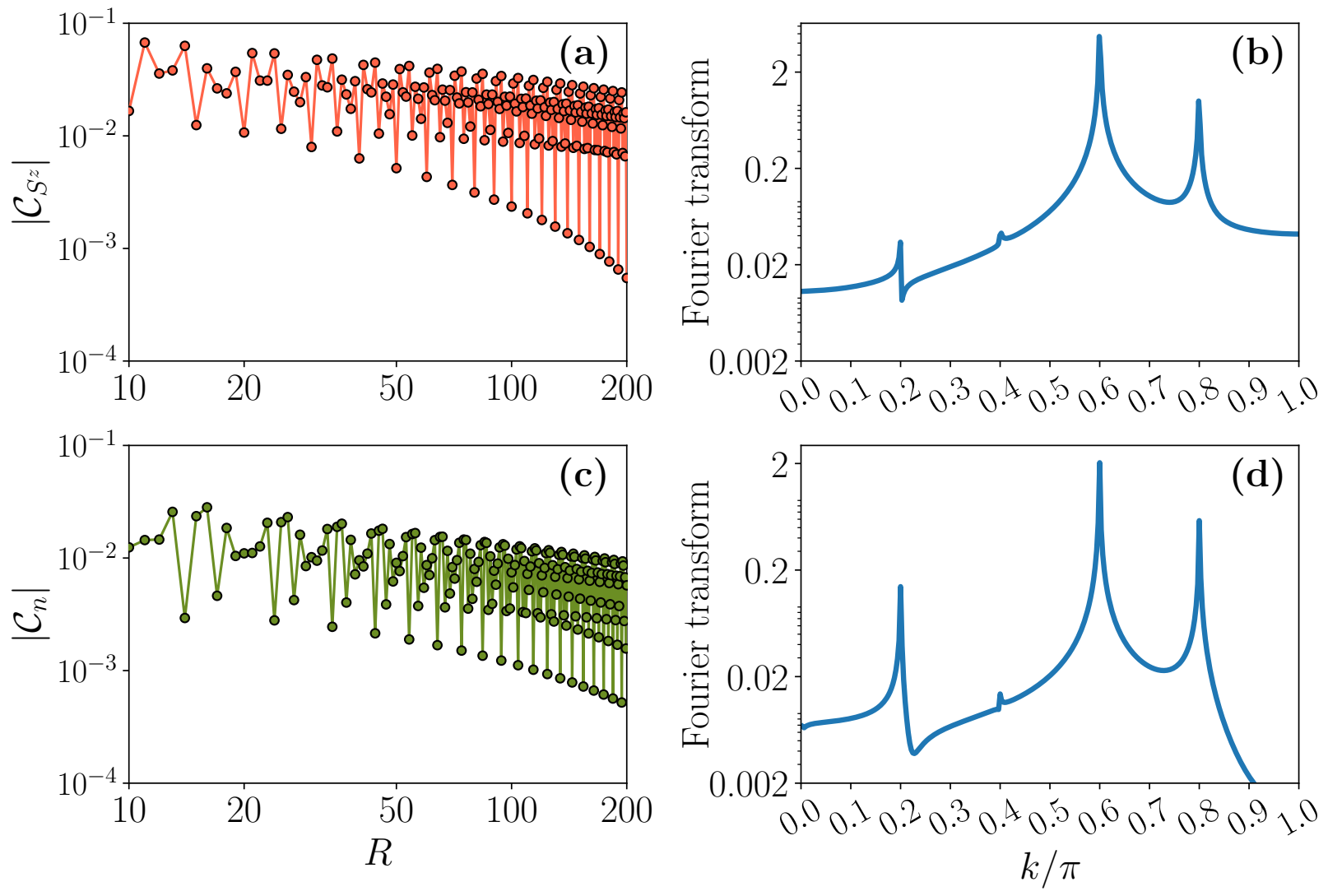

FIG. 5. (Color online.) The behaviors of **(a)** the spin correlation function $\mathcal{C}_{S^z}(R) = \langle S_j^z S_{j+R}^z \rangle - \langle S_j^z \rangle \langle S_{j+R}^z \rangle$ and **(c)** the charge correlation function $\mathcal{C}_n(R) = \langle n_j n_{j+R} \rangle - \langle n_j \rangle \langle n_{j+R} \rangle$ as functions of the distance $R$ in the CLL phase ($U/t = 1, V/t = 5$). The subplots **(b)** and **(d)** depict the Fourier transform of the spin correlation and charge correlation functions respectively.

$\mathcal{C}_{B^C}$ decays exponentially with the distance $R$. The opposite is observed in the CDW phase. The frequencies of the oscillations in the algebraically decaying correlations can be obtained by fitting the the data to the following formula:

$$\mathcal{C}(R) \sim \cos(kR)\, R^{-\beta}. \tag{S.30}$$

Figs. 4**(c)-(d)** show such fits for the bond-spin and the bond-charge correlations respectively in the SDW and CDW phases. In both the cases, the numerical fits show that the frequency of oscillations are $k = 2k_F = 2\pi/5$ as expected from the analysis of Sec. II.

In the case of the CLL phase, since the leading instabilities are clustered, we look at the spin correlation $\mathcal{C}_{S^z}$ and charge correlation $\mathcal{C}_n$ functions. Fig. 5 shows the evolution of these correlation functions with respect to the distance $R$ together with their Fourier transform. As expected in the CLL phase, both the correlation functions decay following a power-law. By analyzing the Fourier transforms, we extract the two primary frequencies of the oscillations as $k = 2k_F = 4\pi/5$ and $k = 3k_F/2 = 3\pi/5$. $k = 3k_F/2$ is the cluster Fourier momentum appearing in Eq. (S.18) that translates the specific cluster instability of the Luttinger liquid. Like in the square ladder [9], we interpret the other major frequencies such as $k = 2k_F$ as due to the pollution of the signal by subdominant SDW or CDW instabilities. Interestingly and unlike the square geometry, these frequencies are present in both spin and charge correlation functions, translating the deformation due to the added range of $U$.

## V. STRONG $U$ COUPLING THEORY IN THE TRIANGULAR LADDER.

We present here the details of the semi-classical approach to the holonic polarized and the cluster crystal phases.

### A. Densities of holons and domain walls

It is possible to get formulae for both the domain walls and holon densities. Because holons can form cohorts, these densities are non-local operators preventing the formulae from being easily used. To obtain them, we define the projectors:

$$P_{i-} = n_{i-}(1 - n_{i+}), \tag{S.31a}$$
$$P_{i+} = n_{i+}(1 - n_{i+1\,+}), \tag{S.31b}$$
$$M_{i-} = (1 - n_{i-})(1 - n_{i+}), \tag{S.31c}$$
$$M_{i+} = (1 - n_{i+})(1 - n_{i+1\,-}), \tag{S.31d}$$
$$D_{i-} = (1 - n_{i-})n_{i+}, \tag{S.31e}$$
$$D_{i+} = (1 - n_{i+})n_{i+1\,-}. \tag{S.31f}$$

Using Eqs. (S.31), we define the quasi-local projector $\mathcal{P}_i$ and the total projector $\mathcal{P}$ onto the effective Hilbert space of strong $U$ coupling as

$$\mathcal{P}_i = (M_{i-1\,+} + D_{i-1\,+} + P_{i-1\,+})(M_{i\,+} + D_{i\,+} + P_{i\,+}) - P_{i-1\,+}P_{i\,+}, \tag{S.32a}$$
$$\mathcal{P} = \otimes_i \mathcal{P}_i. \tag{S.32b}$$

Using Eqs. (S.31), we obtain the density of domain walls by summing over the projectors on one domain wall state for all cohorts. We note $w_i^{/}$ (resp. $w_i^{\backslash}$) a domain wall with its first empty site on rung $i$ and separating a domain with the leg + populated on the left (resp. right) and the leg − populated on the right (resp. left). Their respective density operator are

$$\begin{aligned} n_{wi}^{/} &= P_{i-1\,+}D_{i\,+} + P_{i-1\,+}M_{i+}D_{i+1\,+} + \dots \\ &= P_{i-1\,+}\left(D_{i\,+} + \sum_{k=0}^{L-4} M_{i+}M_{i+1\,+}\dots M_{i+k\,+}D_{i+k+1\,+}\right), \end{aligned} \tag{S.33a}$$

$$\begin{aligned} n_{wi}^{\backslash} &= P_{i-}D_{i+1\,-} + P_{i-}M_{i+1-}D_{i+2\,-} + \dots \\ &= P_{i-}\left(D_{i+1\,-} + \sum_{k=0}^{L-5} M_{i+1\,-}M_{i+2\,-}\dots M_{i+k+1\,-}D_{i+k+2\,-}\right), \end{aligned} \tag{S.33b}$$

for periodic boundary conditions. We also define the total density of domain walls

$$n_{w,i} = n_{wi}^{/} + n_{wi-1}^{\backslash}. \tag{S.34}$$

We obtain the holon densities in a similar manner. We define the two quantities

$$n_{hi+} = P_{i-1+}(1-n_{i+})D_{i+1-} + P_{i-1+}(1-n_{i+})M_{i+1-}D_{i+2-} + P_{i-2+}M_{i-1+}(1-n_{i+})D_{i+1+} + \dots \tag{S.35a}$$
$$+ \frac{1}{2}\left[P_{i-1+}M_{i+}D_{i+1+} + P_{i-1-}M_{i-}D_{i+1-} + \dots\right],$$
$$= \sum_{k=0}^{\infty}\sum_{l=0}^{k} P_{i-1-l+}\prod_{m=1}^{l} M_{i-m+}(1-n_{i+})\prod_{r=1}^{k-l} M_{i+r-}D_{i+(k-l)+1-} \tag{S.35b}$$
$$+ \frac{1}{2}\left[\sum_{k=0}^{\infty}\sum_{l=0}^{k} P_{i-1-l+}\prod_{m=0}^{k} M_{i-l+m+}D_{i+(k-l)+1+} + \sum_{k=0}^{\infty}\sum_{l=0}^{k} P_{i-1-l-}\prod_{m=0}^{k} M_{i-l+m-}D_{i+(k-l)+1-}\right],$$
$$n_{hi-} = \sum_{k=0}^{\infty}\sum_{l=0}^{k} P_{i-1-l-}\prod_{m=1}^{l} M_{i-m+}(1-n_{i-})\prod_{r=1}^{k-l} M_{i+r-1+}D_{i+(k-l)+} \tag{S.35c}$$
$$+ \frac{1}{2}\left[\sum_{k=0}^{\infty}\sum_{l=0}^{k} P_{i-1-l-}\prod_{m=0}^{k} M_{i-l+m-}D_{i+(k-l)+1-} + \sum_{k=0}^{\infty}\sum_{l=0}^{k} P_{i-2-l+}\prod_{m=0}^{k} M_{i-l+m-1+}D_{i+(k-l)+}\right].$$

that would correspond to the densities of holons in a domain polarized on leg + and − respectively. When the holon belongs to the cohort of a domain wall, there is an ambiguity in these definition. This ambiguity explains the factor 1/2 (chosen conventionally). This ambiguity is lifted when defining holons independently of the leg (as it should) *e.g.* for $n_{h,i} = n_{hi+} + n_{hi-}$ when the holon belongs to the cohort of a domain wall $\backslash$ or no cohort of a domain wall and $n_{h,i} = n_{hi-1+} + n_{hi-}$ otherwise. The holon density is then

$$\begin{aligned} n_{hi} = &\sum_{k=0}^{\infty}\sum_{l=0}^{k} P_{i-1-l+}\prod_{m=1}^{l} M_{i-m+}(1-n_{i+})\prod_{r=1}^{k-l} M_{i+r-}D_{i+(k-l)+1-} \\ &+ \sum_{k=0}^{\infty}\sum_{l=0}^{k} P_{i-1-l-}\prod_{m=1}^{l} M_{i-m+}(1-n_{i-})\prod_{r=1}^{k-l} M_{i+r-1+}D_{i+(k-l)+} \\ &+ \sum_{k=0}^{\infty}\sum_{l=0}^{k} P_{i-1-l+}\prod_{m=0}^{k} M_{i-l+m+}D_{i+(k-l)+1+} + \sum_{k=0}^{\infty}\sum_{l=0}^{k} P_{i-2-l-}\prod_{m=0}^{k} M_{i-l+m-1-}D_{i+(k-l)-}. \end{aligned} \tag{S.36}$$

From the densities Eqs. (S.33) and (S.36), we deduce the existence of the (non-local) creation and annihilation operator for both holons ($h_i^\dagger$ and $h_i$) and domain walls ($w_i^\dagger$ and $w_i$).

### B. Estimation of the energy level

We rewrite the perturbative Hamiltonian Eq. (4) using $h_i^\dagger$, $h_i$, $w_i^\dagger$, and $w_i$ to estimate the energy levels as a function of the average densities $f_h$ and $f_w$. The perturbative Hamiltonian is

$$\begin{aligned} H = t\sum_{i=1}^{L}\Big[&h_i^\dagger h_{i+1}\left(1-n_{w^\backslash,i}\right) + h_i^\dagger w_{i+1}^{\backslash\dagger} h_{i+1} w_i^{\backslash} + h_i^\dagger h_{i+1}\left(1-n_{w^/,i+1}\right) + h_i^\dagger w_{i+2}^{/\dagger} h_{i+2} w_{i+1}^{/} \\ &+ h_{i+1}^\dagger h_i\left(1-n_{w^\backslash,i+1}\right) + w_i^{\backslash\dagger} h_{i+1}^\dagger w_{i+1}^{\backslash} h_i + h_{i+1}^\dagger h_i\left(1-n_{w^/,i+2}\right) + w_{i+1}^{/\dagger} h_{i+2}^\dagger w_{i+2}^{/} h_i\Big] \\ &- \frac{2t^2}{U}\sum_i\left[n_{w^\backslash,i}(1-n_{h,i+1}) + n_{w^/,i}(1-n_{h,i+1})\right], \end{aligned} \tag{S.37}$$

leading to the phenomenological Eq. (6) when simplifying the details concerning the range that have no influence on the estimation. For simplicity, we consider holons independent from domain walls to obtain an effective Hamiltonian for each by averaging over the holons or the domain walls separately. We have

$$\begin{aligned} H_h \simeq |\alpha| \sum_{i=1}^{L} \left(e^{i\theta} h_i^\dagger h_{i+1} + \text{H.c.}\right) \\ + \beta \sum_i (n_{h,i} - 1), \end{aligned} \tag{S.38a}$$

$$\begin{aligned} H_w \simeq |\gamma| \sum_{i=1}^{L} \left(e^{i\phi} w_i^\dagger w_{i+1} + \text{H.c.}\right) \\ + \delta \sum_i n_{w,i} + 2tL\text{Re}(\gamma), \end{aligned} \tag{S.38b}$$

with

$$\begin{aligned} \alpha &= t\left(\langle 1 - n_{w,i+1}\rangle + \langle w_{i+1}^\dagger w_i\rangle\right), \\ &= |\alpha| e^{i\theta}, \end{aligned} \tag{S.39a}$$

$$\beta = \frac{2t^2}{U}\langle n_{w,i}\rangle, \tag{S.39b}$$

$$\begin{aligned} \gamma &= t\langle h_{i+1}^\dagger h_i\rangle, \\ &= |\gamma| e^{i\phi}, \end{aligned} \tag{S.39c}$$

$$\begin{aligned} \delta = &-t\left(\langle h_{i-1}^\dagger h_{i-2}\rangle + \langle h_{i-1}^\dagger h_i\rangle\right) \\ &- \frac{2t^2}{U}\langle 1 - n_{h,i}\rangle. \end{aligned} \tag{S.39d}$$

By diagonalizing the Hamiltonian Eqs. (S.38) and building the eigenstates as a product state of holonic and domain wall mode, we get the hopping averages (*e.g.* $\langle w_{i+1}^\dagger w_i\rangle$) such that

$$\alpha = t\left(1 - f_s\right) + \frac{2t}{L} e^{i(\phi - \pi - \frac{\pi}{L})} \frac{\sin\left(\frac{\pi}{2} f_w\right)}{\sin\left(\frac{\pi}{L}\right)}, \tag{S.40a}$$

$$\beta = \frac{2t^2 f_w}{U}, \tag{S.40b}$$

$$\gamma = \frac{t}{L} e^{i(\theta - \pi - \frac{\pi}{L})} \frac{\sin(\pi f_h)}{\sin\left(\frac{\pi}{L}\right)}, \tag{S.40c}$$

$$\begin{aligned} \delta = &-\frac{2t}{L}\cos\left(\theta - \pi - \frac{\pi}{L}\right)\frac{\sin(\pi f_h)}{\sin\left(\frac{\pi}{L}\right)} \\ &- \frac{2t^2}{U}\left(1 - f_h\right), \end{aligned} \tag{S.40d}$$

$$\phi = \theta - \pi - \pi/L \ [2\pi], \tag{S.40e}$$

$$\tan(\theta) = \frac{\sin\left(\theta - \frac{2\pi}{L}\right)}{\cos\left(\theta - \frac{2\pi}{L}\right) + \frac{L\sin(\pi/L)}{\sin(\pi f_w)}\left(1 - f_w\right)}. \tag{S.40f}$$

Eqs. (S.40e) and (S.40f) always (only) admit $\theta = \phi + \pi = 0\ [2\pi]$ as a solution when $L \to \infty$. We then obtain Eq. (7) by averaging Eq. (6) over both holons and domain walls successively and using Eqs. (S.40) in the limit $L \to \infty$.

### C. Phenomenological shoulder potential

We provide the expression of the shoulder potential and its contribution to the estimate of the energy level for $r_c = 3$. The method is identical to Sec. V B, but must be carried term by term. It is diagonal in the basis of domain walls and holons so there are no higher order corrections in $V$. Starting with

$$H_{\text{Shoulder}} = \sum_{i,\ell} \sum_{r=1}^{r_c} V_r n_{i,\ell} n_{i+r,\ell}, \tag{S.41}$$

the shoulder potential is rewritten as

$$
\begin{aligned}
H_{\text{Shoulder}} - L(V_1+V_2+V_3) = &-(V_1+2V_2+3V_3)\sum_i n^{\backslash}_{w,i} - (2V_1+3V_2+4V_3)\sum_i n^{/}_{w,i} \\
&- 2(V_1+V_2+V_3)\sum_i n_{h,i} + V_1\sum_i n_{h,i}n_{h,i+1} + V_2\sum_i n_{h,i}n_{h,i+2} + V_3\sum_i n_{h,i}n_{h,i+3} \\
&+ (V_1+V_2+V_3)\sum_i n^{/}_{w,i}n_{h,i+1} + (V_1+V_2+V_3)\sum_i n^{\backslash}_{w,i}n_{h,i+1} + (V_2+V_3)\sum_i n^{/}_{w,i}n_{h,i+2} \\
&+ (V_2+V_3)\sum_i n^{\backslash}_{w,i}n_{h,i+2} + (V_2+V_3)\sum_i n_{h,i}n^{/}_{w,i+2} + (V_2+V_3)\sum_i n_{h,i}n^{\backslash}_{w,i+1} \\
&+ V_3\sum_i n^{/}_{w,i}n_{h,i+3} + V_3\sum_i n^{\backslash}_{w,i}n_{h,i+3} + V_3\sum_i n_{h,i}n^{/}_{w,i+3} + V_3\sum_i n_{h,i}n^{\backslash}_{w,i+2} \\
&+ (V_2+2V_3)\sum_i (n^{/}_{w,i}n^{\backslash}_{w,i+1} + n^{\backslash}_{w,i}n^{/}_{w,i+2}), + V_3\sum_i (n^{/}_{w,i}n^{\backslash}_{w,i+2} + n^{\backslash}_{w,i}n^{/}_{w,i+3}),
\end{aligned} \tag{S.42a}
$$

$$
\begin{aligned}
\sim -\frac{1}{2}\sum_{i,l}(1+2l)V_l n_{w,i} - 2\sum_{i,l} V_l n_{h,i} + \sum_{i,l} V_l n_{h,i}n_{h,i+l} + 2\sum_{i,l}\sum_{m=l}^{\infty} V_m n_{w,i}n_{h,i\pm l} \\
+ (V_2+2V_3)\sum_i (n^{/}_{w,i}n^{\backslash}_{w,i+1} + n^{\backslash}_{w,i}n^{/}_{w,i+2}) + V_3\sum_i (n^{/}_{w,i}n^{\backslash}_{w,i+2} + n^{\backslash}_{w,i}n^{/}_{w,i+3}),
\end{aligned} \tag{S.42b}
$$

where Eq. (S.42b) is the phenomenological approximation of Eq. (S.42a). The contribution of the shoulder potential to the energy levels follows:

$$
\begin{aligned}
\langle H_{\text{Shoulder}} \rangle = L\sum_l V_l - \frac{L}{2}\sum_l (1+2l)V_l f_w - 2L\sum_l V_l f_h + L\sum_l V_l\left(f_h^2 - \frac{1}{l\pi^2}\sin(\pi f_h)\sin(l\pi f_h)\right) \\
+ L\sum_l (2l-1)V_l f_m f_w + L(V_2+3V_3)f_w^2/2.
\end{aligned} \tag{S.43}
$$

When $V_1 = V_2 = V$ and $V_3 = 0$, we find Eq. (8).

---